\documentclass[12pt]{article}

\usepackage[totalheight = 23cm, totalwidth = 17cm]{geometry}
\usepackage{amssymb,amsmath,amsfonts,amsbsy,epsfig}
\usepackage{bm}

\newcommand{\mpl}{m_{\rm Pl}}

\begin{document}

\begin{titlepage}

\begin{center}

\vspace*{-10ex}
\hspace*{\fill}
YITP-08-30

\vskip 1.5cm

\Huge{Curvature perturbation spectrum
\\
from false vacuum inflation}

\vskip 1cm

\large{
Jinn-Ouk Gong$^{1}$\footnote{jgong\_AT\_ hep.wisc.edu}
\hspace{0.2cm}\mbox{and}\hspace{0.2cm}
Misao Sasaki$^2$\footnote{misao\_AT\_yukawa.kyoto-u.ac.jp}
\\
\vspace{0.5cm}
{\em
${}^1$ Department of Physics, University of Wisconsin-Madison
\\
1150 University Avenue, Madison, WI 53706-1390, USA
\\
\vspace{0.2cm}
${}^2$ Yukawa Institute for Theoretical Physics
\\
Kyoto University, Kyoto 606-8502, Japan}
}

\vskip 0.5cm

\today

\vskip 1.2cm

\end{center}

\begin{abstract}

In the inflationary cosmology it occurs frequently that the inflaton field is
trapped in a local, transient minimum with non-zero vacuum energy. The difficulty
regarding the curvature perturbation produced during such a stage is that
classically the inflaton does not move so that the comoving hypersurfaces are not
well defined at linear order in the scalar field perturbation. In this paper,
assuming a mechanism of trapping which resembles a high temperature correction to
the potential, we explicitly calculate for the first time the resulting power
spectrum of the curvature perturbation by evaluating the quantum two-point
correlation function directly. The spectrum is steeply blue with the spectral index
$n_\mathcal{R} = 4$.

\end{abstract}

\end{titlepage}

\setcounter{page}{0}
\newpage
\setcounter{page}{1}

\section{Introduction}

Now it is widely believed that inflation~\cite{Guth:1980zm,inf} takes place at the
earliest moments in the history of the universe and that after inflation the initial
conditions are all satisfied necessary for the successful hot big bang universe. One
of the greatest triumphs of inflation is that we can naturally derive a nearly scale
invariant spectrum $\mathcal{P}_\mathcal{R}$ of the comoving curvature perturbation
$\mathcal{R}_c$~\cite{Liddle:2000cg} which is required by the recent observations,
including the Wilkinson Microwave Anisotropy Probe 5-year data~\cite{Komatsu:2008hk}
where $n_\mathcal{R} \approx 0.96$. This nearly scale invariant spectrum is
generated under the slow-roll approximation where the inflaton field $\phi$ is very
slowly evolving towards the global minimum of its effective potential. The
calculation of $\mathcal{P}_\mathcal{R}$ is now a well established
subject~\cite{Kodama:1985bj,Mukhanov:1990me} and has become refined with high
accuracy~\cite{higherordercalc}.

However, the slow-roll phase is not a necessary condition for inflation and thus the
inflationary prediction of a nearly scale invariant $\mathcal{P}_\mathcal{R}$ is not
necessarily true. A typical situation where $\phi$ is not slowly rolling off the
potential is when it is trapped in a local minimum, i.e. false vacuum: for example
in the original scenario of inflation~\cite{Guth:1980zm} it is assumed that $\phi$
is confined in a local minimum and the inflationary epoch ends by quantum tunneling.
Moreover, although this original scenario is observationally not viable, such a
phase needs not be ruled out a priori. It may lie far outside the observable regime
of the last 60 $e$-folds of expansion out of the whole period of
inflation\footnote{Note that in the so-called locked inflation~\cite{Dvali:2003vv},
most of the observationally relevant part of the universe exits the horizon when
$\phi$ is effectively trapped in a transient local minimum. However, the stable
false vacuum is supported by the rapid oscillation of another scalar field coupled
to $\phi$. This is different from what we are going to discuss.}. Also, in the case
of thermal inflation~\cite{thermalinflation} a short period of inflation is provided
by a constant vacuum energy due to a temperature effect at the end of conventional
inflation.

A problematic fact is that, for such a period the standard calculation of
$\mathcal{R}_c$ does not work. The reason is that when $\phi$ is trapped in a false
vacuum, classically $\dot\phi = 0$\footnote{As there is {\em no} background
evolution of $\phi$, it may not be quite proper to call it `inflaton'. Nevertheless,
we call it the inflaton field simply because its potential energy at the false
vacuum is the cause of the inflationary de Sitter expansion.}, so that the comoving
curvature perturbation, which is given by
\begin{equation}
\mathcal{R}_c \sim \frac{H}{\dot\phi}\delta\phi \, ,
\end{equation}
where $H = \dot{a}/a$ is the Hubble parameter, is not defined. The form of
$\mathcal{P}_\mathcal{R}$ has been roughly guessed~\cite{Gong:2006hf} but its exact
functional form has not yet been known\footnote{Note that in Ref.~\cite{Pilo:2004ke}
the spectrum of the field fluctuations, $\mathcal{P}_{\delta\phi}$, is calculated.}.
But this never means that the situation itself is singular, but that we need to
adopt a different way of calculation to obtain $\mathcal{P}_\mathcal{R}$: we should
derive the final result without resorting to the classical homogeneous scalar field
background~\cite{Nambu:1989eu}.

In this paper, for the first time we explicitly calculate $\mathcal{P}_\mathcal{R}$
and the corresponding spectral index, $n_\mathcal{R}$, from a stage of false vacuum
inflation. What is important is that the perfect de Sitter phase does not last
forever. It should eventually end. There are a number of ways to terminate this pure
de Sitter expansion. Here we adopt a mechanism like thermal inflation. To be
specific, we consider an effective mass-squared which consists of a negative
constant term corresponding to a bare mass-squared term and a positive term
proportional to $a^{-2}$. The latter term is equivalent to a temperature effect
$g^2T^2\propto a^{-2}$~\cite{Linde:1982uu}, i.e. such a term can arise due to
possible couplings to thermal bath.

This paper is outlined as follows. In Section~\ref{section_2pointcorrfct}, we write
the two-point correlation functions of the inflaton and the energy density. In
Section~\ref{section_powerspectra} we first calculate the power spectrum of the
gauge invariant intrinsic spatial curvature perturbation $\Phi$ using the two-point
correlation function calculated in the previous section. Then we extract the final
form of $\mathcal{P}_\mathcal{R}$. Finally, we conclude in
Section~\ref{section_conclusions}.

\section{Two-point correlation functions}
\label{section_2pointcorrfct}

\subsection{Inflaton field two-point correlation function}

We consider a theory with the Einstein-scalar Lagrangian,
\begin{eqnarray}
L=\frac{m_{\rm Pl}^2}{2}R-\frac{1}{2}g^{\mu\nu}\partial_\mu\phi\partial_\nu\phi
-V(\phi;t)\,,
\label{lagrangean}
\end{eqnarray}
where $\mpl^{-2} \equiv 8\pi G$ and the potential is assumed to have the form,
\begin{eqnarray}
V=V_0+\frac{1}{2}m_{\rm eff}^2\phi^2\,,
\label{potform}
\end{eqnarray}
where
\begin{eqnarray}
m_{\rm eff}^2=m_\phi^2+\frac{\mu^2}{a^2}\,,
\label{mass}
\end{eqnarray}
with $V_0>0$, $m_\phi^2<0$ and $\mu^2>0$. We consider the stage when the effective
mass-squared is positive, $m_{\rm eff}^2>0$, so that the inflaton is classically
trapped at $\phi=0$. The background Hubble parameter is given by
\begin{eqnarray}
3H^2=\frac{V_0}{m_{\rm Pl}^2}\,,
\end{eqnarray}
and the cosmic scale factor during this stage can be well approximated by the pure
de Sitter expression,
\begin{eqnarray}
a=a_*\exp[H(t-t_*)]=\frac{1}{-H\eta}\,,
\label{scalefactor}
\end{eqnarray}
where $t_*$ is an arbitrary fiducial time and $\eta$ is the conformal time. Note
that a condition for $\phi=0$ to be sufficiently stable is $m_{\rm eff}^2/H^2\gg1$,
hence we must have
\begin{eqnarray}
\frac{\mu^2}{H^2a^2}=\mu^2\eta^2\gg 1\,.
\label{timerange}
\end{eqnarray}
In the following we focus on this stage. We also note that it is generally assumed
that $|m_\phi^2|>H^2$ in the case of thermal inflation.

Since $\phi$ is trapped in a false vacuum so that $\langle\phi\rangle=0$ during
this stage of pure de Sitter expansion, the scalar field perturbation is
in fact equal to the scalar field itself, i.e.
\begin{equation}\label{deltaphiasphi}
\delta\phi = \phi - \langle\phi\rangle = \phi \, .
\end{equation}
Now we begin with considering the two-point function
\begin{align}
G(x,x') & = \langle \phi(x)\phi(x') \rangle
\nonumber \\
& = \frac{1}{2} \langle \phi(x)\phi(x') + \phi(x')\phi(x) \rangle
\nonumber \\
& = \frac{1}{2} G^{(1)}(x,x') \, ,
\end{align}
where $G^{(1)}(x,x')$ is the symmetric two-point function. We have an {\em exact\/}
expression for $G^{(1)}(x,x')$ as~\cite{Bunch:1978yq}
\begin{align}
\label{preG}
G^{(1)}(x,x') =& \frac{H^2}{2\pi^2} \int_0^\infty ds \cosh(\nu s) \frac{1 + p
(2\cosh{s} - 2Z)^{1/2}}{(2\cosh{s} - 2Z)^{3/2}}
\nonumber \\
&\hspace{1.5cm} \times \exp\left[ -p (2\cosh{s} - 2Z)^{1/2} \right] \, ,
\end{align}
where
\begin{align}
p = & \sqrt{\mu^2\eta\eta'} \, ,
\\
Z = & \frac{\eta^2 + {\eta'}^2 - r^2}{2\eta\eta'} \, ,
\\
r^2 = & |\bm{x} - \bm{x}'|^2 \,,
\end{align}
and
\begin{equation}\label{nu}
\nu^2 = \frac{9}{4} - \frac{m_\phi^2}{H^2} > \frac{9}{4} \, .
\end{equation}
Note that the form of $G^{(1)}(x,x')$ of our interest is the one in the limit $p \gg
1$, or equivalently in the early stage of inflation $\eta \to -\infty$\footnote{Note
that the opposite limit $\eta\to 0$ is discussed in Ref.~\cite{Nambu:1989eu} and the
spectrum is consistent with the standard result $\mathcal{P}_\mathcal{R} =
[H^2/(2\pi\dot\phi)]^2$.}, when $\phi$ is trapped in a transient local minimum.

A technically important point is the existence of the term $\exp\left[ -p (2\cosh{s}
-2Z)^{1/2} \right]$ in Eq.~(\ref{preG}). Since $\cosh{x}$ is exponentially
increasing as $x$ increases and we are interested in the limit $p \gg 1$, this term
is highly suppressed for large $s$, making contribution to the integral from this
region negligible. Thus, the dominant contribution of the integral comes from the
region near $s \approx 0$, and hence we can expand the hyperbolic cosine function
around this region and take only the leading term. Using $\cosh{x} = 1 + x^2/2 +
\cdots$, we have
\begin{align}
G^{(1)}(x,x') & \approx \frac{H^2}{2\pi^2} \int_0^\infty ds \left( 1 +
\frac{\nu^2s^2}{2} \right)
\frac{1+ p \left[ 2 (1 + s^2/2) - 2Z \right]^{1/2}}{[2(1 + s^2/2) - 2Z]^{3/2}}
\nonumber \\
& \hspace{1.5cm} \times \exp \left\{ -p \left[2 \left( 1 + \frac{s^2}{2} \right)
 -2Z \right]^{1/2} \right\} \, .
\end{align}
Another point to be kept in mind is that we are ultimately interested only in
super-horizon scales, just as in the case of standard slow-roll inflation. That is,
the two points $x$ and $x'$ are space-like separated with their distance being much
larger than $H^{-1}$. Thus introducing a new variable
\begin{equation}
u \equiv 1 - Z = \frac{r^2 - (\eta - \eta')^2}{2\eta\eta'} \, ,
\end{equation}
we find for $u\gg1$,
\begin{align}
G^{(1)}(x,x') & \approx \frac{H^2}{2\pi^2} \frac{p}{2u}
\int_0^\infty ds \exp\left[-p (s^2 + 2u)^{1/2} \right]
\nonumber \\
& = \frac{H^2}{2\pi^2} \frac{p}{\sqrt{2u}} K_1(p\sqrt{2u}) \, ,
\end{align}
where we have used an identity of the modified Bessel function of the second kind
$K_\nu(x)$,
\begin{equation}
\int_0^\infty e^{-x \sqrt{t^2 + z^2}} dt = zK_1(xz) \, .
\end{equation}
Now, using the asymptotic form
\begin{equation}
K_\nu(z) \underset{z \gg 1}{\longrightarrow} \sqrt{\frac{\pi}{2z}}e^{-z} \, ,
\end{equation}
we have
\begin{equation}
G^{(1)}(x,x') \approx \frac{H^2}{2\pi^2} \sqrt{\frac{\pi}{2}}
\frac{\sqrt{p}}{(2u)^{3/4}} e^{-p\sqrt{2u}} \, .
\end{equation}
Therefore, the two-point function in the regime of our interest
is given by
\begin{align}\label{G}
G(x,x') & = \frac{1}{2}G^{(1)}(x,x')
\nonumber \\
& \approx \left( \frac{H}{2\pi} \right)^2 \sqrt{\frac{\pi}{2}}
\frac{\sqrt{\mu}\eta\eta'}{\left[ r^2 - (\eta - \eta')^2 \right]^{3/4}} \exp
\left\{-\mu \left[r^2 - (\eta - \eta')^2\right]^{1/2} \right\} \, .
\end{align}

\subsection{Energy density two-point correlation function}

Although there is no classically evolving background scalar field at the stage of
our interest, this does not mean that there is no energy density fluctuations. In
fact, because of the quantum vacuum fluctuations of the inflaton field, there exist
fluctuations in its energy-momentum tensor. To evaluate the curvature perturbation
from this stage, calculating the energy-momentum tensor in the pure de Sitter
background is not sufficient. We have to take into account the metric perturbation.
 But under the situation of our interest where there is no background
evolution of $\phi$, it is exactly the same as the standard quantum field theory in
curved space-time. Thus there is no metric perturbation at linear order in the field
fluctuations $\delta\phi$, or the scalar field itself $\phi$: see
Eq.~(\ref{deltaphiasphi}). So in the previous section it is perfectly legitimate to
consider $\phi$ in the given homogeneous and isotropic background. The metric
perturbation $\delta{g}_{\mu\nu}$ appears at second order in $\phi$, i.e. it is
linear in the perturbation of the energy-momentum tensor. Then we can apply the
standard linear perturbation theory. This is what we are going to do in this
section.

The energy-momentum tensor is given by
\begin{eqnarray}
T_{\mu\nu}=\partial_\mu\phi\partial_\nu\phi
-\frac{1}{2}g_{\mu\nu}
\left(g^{\alpha\beta}\partial_\alpha\phi\partial_\beta\phi
+2V\right)\,,
\label{Tmn}
\end{eqnarray}
and we set
\begin{eqnarray}
\delta T^{\mu}{}_{\nu} =T^\mu{}_\nu-\left\langle T^\mu{}_\nu \right\rangle\,.
\label{Tmnfluc}
\end{eqnarray}
It is important to note that, as can be read from Eqs.~(\ref{Tmn}) and
(\ref{Tmnfluc}), the linearly perturbed energy-momentum tensor $\delta{T}_{\mu\nu}$
is quadratic in $\phi$, at which order the metric perturbation comes into play and
we can follow the standard cosmological perturbation theory: the gauge-invariant
density perturbation $\Delta$, which is the density perturbation on the comoving
hypersurface on which $T^0{}_i=0$, is expressed as~\cite{Kodama:1985bj,Nambu:1989eu}
\begin{eqnarray}
\nabla^2(\rho\Delta)
=\nabla^2(-T^0{}_0)+3H\partial^i(-T^0{}_i)\,,
\label{Delta}
\end{eqnarray}
where $\nabla^2=\delta^{ij}\partial_i\partial_j$ and we have chosen the time
coordinate to be the cosmic proper time; $x^0=t$.

Now we introduce the two-point correlation function of $\nabla^2(\rho\Delta)$,
\begin{align}\label{corrD}
D(x,x') & \equiv \left\langle \nabla_x^2 \left[ \rho\Delta(x) \right] \nabla_{x'}^2
\left[ \rho\Delta(x') \right] \right\rangle
\nonumber \\
& = f_i^{\rho\mu\nu}(t) f_{j'}^{\sigma'\alpha'\beta'}(t) \partial^i\partial^{j'}
\left\{ \left[ \partial_\rho\partial_{\alpha'}\partial_{\beta'}G(x,x') \right]
\left[ \partial_{\sigma'}\partial_\mu\partial_\nu G(x,x') \right] \right.
\nonumber \\
& \hspace{4.5cm} \left. + \left[ \partial_\rho\partial_{\sigma'} G(x,x') \right]
\left[ \partial_\mu\partial_\nu\partial_{\alpha'}\partial_{\beta'} G(x,x') \right]
\right\} \, .
\end{align}
Using the expression for the energy-momentum tensor, Eq.~(\ref{Tmn}), the
coefficients $f_i^{\rho\mu\nu}$ are found to be~\cite{Nambu:1989eu}
\begin{align}
\label{coeff1}
f_i^{00j} = f_i^{0j0} = & \frac{1}{2}\delta_i^{\,\,j} \, ,
\\\label{coeff2}
f_i^{j00} = & -\delta_i^{\,\,j} \, ,
\\\label{coeff3}
f_i^{jkl} = & a^{-2} \left[ \delta_i^{\,\,j}\delta^{kl} + \frac{1}{2} \left(
\delta_i^{\,\,k}\delta^{jl} + \delta_i^{\,\,l}\delta^{jk} \right) \right] \, ,
\end{align}
and zero otherwise, where the potential dependence is eliminated in favor of the
spacetime derivatives using the field equation for the scalar field. Then
substituting the coefficients into Eq.~(\ref{corrD}), collecting non-zero
components, expanding the Kronecker delta terms and finally rearranging the indices,
we obtain a rather lengthy expression given in Appendix~\ref{2pointfct_formula},
Eq.~(\ref{expD}). There the time derivatives are those with respect to $t$. For
later purpose, it is convenient to express the time dependence in terms of the
conformal time $\eta$ defined by $d\eta=dt/a$. Since the two-point function depends
only on the comoving distance between the two points $r=|\bm{x}-\bm{x}'|$, we may
then express the coordinate dependence of $D$ as
\begin{eqnarray}
D=D(r;\, \eta,\eta')=D(r;\,\eta',\eta)\,,
\end{eqnarray}
where the second equality comes from the fact that $D$ is symmetric under the
exchange of $\eta$ and $\eta'$.

Now, inserting Eq.~(\ref{G}) into Eq.~(\ref{expD}), we find after some amount of
calculations that, interestingly, the most significant contribution comes from the
terms without any time derivatives, i.e. the terms multiplied by $a^{-4}$ in
Eq.~(\ref{expD}). To leading order, the two-point correlation function $D(x,x')$
evaluated at an equal time $\eta=\eta'$ is given by
\begin{equation}\label{resultD}
D(x,x') \approx \left(\frac{H}{2\pi}\right)^4 16\pi (H\eta)^4
\frac{(\mu\eta)^4}{(\mu r)^3} \mu^8 e^{-2\mu r} \,.
\end{equation}
We note that we can obtain the same result by substituting the exact expression
Eq.~(\ref{preG}) into Eq.~(\ref{expD}) first and then making use of
Eqs.~(\ref{spderiv}) and (\ref{tderiv}), and finally collecting the leading terms.

\section{Power spectra}
\label{section_powerspectra}

\subsection{Power spectrum of $\Phi_{\bm{k}}$}

With the two-point correlation function given as Eq.~(\ref{resultD}), we next turn
to its Fourier transformation. Denoting the Fourier transformation of a function
$f(r)$ to be
\begin{equation}
\mathcal{F}[f](k) \equiv
 \int \frac{d^3r}{(2\pi)^{3/2}} f(r) e^{-i\bm{k}\cdot\bm{r}} \, ,
\end{equation}
then
\begin{align}
\mathcal{F}[D](k) & \approx \left(\frac{H}{2\pi}\right)^4 16\pi (H\eta)^4
\frac{(\mu\eta)^4}{\mu^3} \mu^8 \int \frac{d^3r}{(2\pi)^{3/2}} \frac{e^{-2\mu
r}}{r^3} e^{-i \bm{k \cdot r}}
\nonumber\\
& = \left(\frac{H}{2\pi}\right)^4 16\pi (H\eta)^4 \frac{(\mu\eta)^4}{\mu^3} \mu^8
\frac{4\pi}{(2\pi)^{3/2}} \int_0^\infty dr \, \frac{e^{-2\mu r}}{r}j_0(kr) \, ,
\end{align}
where the spherical Bessel function $j_0$ is given by
\begin{equation}
j_0(x) = \frac{\sin x}{x} \, .
\end{equation}
Notice that the function $\exp(-2\mu r) j_0(kr)/r$ blows up to infinity at $r = 0$
so the integral does not converge.

An important point to remember at this stage is that, we are interested in the
correlations of two points which are separated by super-horizon scales,
$r\gg|\eta|$. Thus, the singularity at $r=0$ should not matter and we may introduce
a cutoff at a small $r$ for the range of integration. Since $\mu^2\eta^2\gg1$ by
assumption, This implies that the region of our interest satisfies $\mu r \gg 1$.
Hence, a natural choice of this cutoff scale would be $1/\mu$. Also, since we are
interested in very large scales, i.e. very small $k$ regions, we can expand
$\sin(kr) = kr - (kr)^3/3! +\cdots$. Then, with the modified integration range, we
have
\begin{align}\label{intresult}
\int_0^\infty dr \frac{e^{-2\mu r}}{r} j_0(kr) \to \int_{1/\mu}^\infty dr
\frac{e^{-2\mu r}}{r} j_0(kr) =
 & \frac{1}{k} \int_{1/\mu}^\infty dr \frac{e^{-2\mu r}}{r^2} \sin(kr)
\nonumber\\
\approx & \frac{1}{k} \int_{1/\mu}^\infty dr \frac{e^{-2\mu r}}{r^2}
\left( kr - \frac{k^3r^3}{6} \right)
\nonumber\\
= & -{\rm Ei}(-2) - \frac{1}{8e^2}\left(\frac{k}{\mu}\right)^2 \, ,
\end{align}
where we have used the definition of the exponential integral function
\begin{equation}
-{\rm Ei}(-x) \equiv \int_x^\infty \frac{e^{-t}}{t} dt = -\gamma - \log x -
\sum_{n=1}^\infty \frac{(-1)^n x^n}{n \cdot n!} \, ,
\end{equation}
with $\gamma \approx 0.577216$ being the Euler-Mascheroni constant, so $-{\rm
Ei}(-2)\approx 0.0489005 > 0$. It should be noted that the argument $-2$ here is due
to our choice of the lower cutoff of the integration, $r=1/\mu$. In general one may
choose any value for the lower cutoff of $r$ as long as $r=\mathcal{O}(1/\mu)$.
Hence we should not regard the actual value of $-{\rm Ei}(-2)$ to be quantitatively
meaningful. Instead we should regard it as giving a factor of order unity. In any
case, as clear from the above, if we are interested in the range of $k$ such that
$k\ll \mu$, that is, the modes which leave the horizon before the time
$-\eta=1/\mu$, the second term proportional to $(k/\mu)^2$ is negligible.

Now we can explicitly write the power spectrum of $\Phi$, the gauge invariant
intrinsic spatial curvature perturbation in the Newtonian (or longitudinal) gauge.
From the perturbed Einstein equations, we have the well known
relation~\cite{Kodama:1985bj}
\begin{equation}
\frac{\nabla^2}{a^2}\Phi = -4\pi G \rho\Delta \, ,
\end{equation}
where the factor $\rho\Delta$ shows up which appears in the definition of $D(x,x')$,
Eq.~(\ref{corrD}). Hence
\begin{equation}\label{rhodelta}
\rho\Delta = -\frac{1}{4\pi G} \frac{\nabla^2}{a^2}\Phi = -2m_\mathrm{Pl}^2
(H\eta)^2 \nabla^2\Phi \,.
\end{equation}
Since we are interested in the correlation function of two points which are apart on
super-horizon scales, the leading contribution of the spatial gradient on the
function of the form $e^{-2\mu r}/r^3$ gives
\begin{equation}
\nabla^2\left( \frac{e^{-2\mu r}}{r^3} \right) = \frac{1}{r^2}
\frac{\partial}{\partial r} \left[ r^2 \frac{\partial}{\partial r} \left(
\frac{e^{-2\mu r}}{r^3} \right) \right]
\approx (2\mu)^2 \frac{e^{-2\mu r}}{r^3} \,.
\end{equation}
Thus, substituting Eq.~(\ref{rhodelta}) into Eq.~(\ref{corrD}),
\begin{align}\label{DandPhi}
D(x,x') & = 4m_\mathrm{Pl}^4(H\eta)^4 \left\langle \nabla_x^2 \left[ \nabla_x^2
\Phi(x) \right] \nabla_{x'}^2 \left[ \nabla_{x'}^2 \Phi(x') \right] \right\rangle
\nonumber\\
& \approx 4m_\mathrm{Pl}^4(H\eta)^4 (2\mu)^8 \langle \Phi(x) \Phi(x') \rangle \, .
\end{align}
Therefore, equating Eq.~(\ref{resultD}) with Eq.~(\ref{DandPhi}), the two-point
correlation function in {\em configuration space} is given by
\begin{align}
\xi_\Phi(r) \equiv & \langle \Phi(\bm{x}) \Phi(\bm{x + r}) \rangle
\nonumber\\
\approx & \frac{1}{4m_\mathrm{Pl}^4(H\eta)^4 (2\mu)^8} \left(\frac{H}{2\pi}\right)^4
16\pi (H\eta)^4 \frac{(\mu\eta)^4}{(\mu r)^3} \mu^8 e^{-2\mu r}
\nonumber\\
= & \frac{\pi}{64} \left( \frac{H}{2\pi m_\mathrm{Pl}} \right)^4
\frac{(\mu\eta)^4}{(\mu r)^3} e^{-2\mu r} \, ,
\end{align}
where we can see that it is exponentially suppressed. Now taking the inverse Fourier
transformation, we can relate $\xi_\Phi$ and $\mathcal{P}_\Phi$ as
\begin{equation}
\mathcal{P}_\Phi
 = \frac{k^3}{2\pi^2} \int d^3r \xi_\Phi(r) e^{-i\bm{k \cdot r}}\, ,
\end{equation}
which includes the integral $\int_0^\infty e^{-2\mu r}j_0(kr)/r$ that we have
already calculated and is given by Eq.~(\ref{intresult}).
We have thus
\begin{equation}
\int d^3r \xi_\Phi(r) e^{-i\bm{k \cdot r}} \approx -{\rm Ei}(-2) \frac{\pi^2}{16}
\left( \frac{H}{2\pi m_\mathrm{Pl}} \right)^4 \frac{(\mu\eta)^4}{\mu^3} \, .
\end{equation}
Note that we have $\mu|\eta| \gg 1$ so that the temperature term, the second term of
Eq.~(\ref{mass}), dominates the effective mass-squared, and hence $\Phi
\propto\eta^2$. This implies that $\Phi$ we are calculating is decaying in time,
hence it may seem that it does not contribute to the final power spectrum at all.
However, it turns out that this behavior of $\Phi$ correctly corresponds to the
growing adiabatic mode as we shall see later. To summarize, we have the power
spectrum of $\Phi$ to leading order as
\begin{equation}\label{Phispectrum}
\mathcal{P}_\Phi(k;\eta) \approx \frac{-{\rm Ei}(-2)}{32}
\left( \frac{H}{2\pi m_\mathrm{Pl}} \right)^4
 (\mu\eta)^4 \left(\frac{k}{\mu}\right)^3 \, .
\end{equation}
To translate this into the power spectrum of the comoving curvature perturbation
$\mathcal{P}_\mathcal{R}$, we need to know $\langle \rho + p \rangle$ which is
classically 0.

\subsection{Power spectrum of the comoving curvature perturbation}

\subsubsection{$\langle \rho + p \rangle$}

As it is well known, the adiabatic density perturbations responsible for the large
scale structure of the universe today is represented by the curvature perturbation
on comoving hypersurfaces $\mathcal{R}_c$ on super-horizon scales. However, the
definition of the comoving hypersurface, $T^0{}_i=0$, becomes meaningless in the
pure de Sitter space. This is simply because $T_{\mu\nu}\propto g_{\mu\nu}$ in pure
de Sitter space, hence $T^0{}_i$ is identically zero. In other words, there exists
no preferred rest frame in pure de Sitter space. Nevertheless, in the present case,
we do have a preferred frame because of the time dependence of $m_{\rm eff}^2$.
Therefore, the vacuum expectation value of $T_{\mu\nu}$ will no longer be de Sitter
invariant. In particular, we expect $\rho+p$ to have a small but non-vanishing
vacuum expectation value. This fact enables us define the comoving hypersurface and
hence the spectrum of the comoving curvature perturbation,
$\mathcal{P}_\mathcal{R}$. Here we assume that the thermal contribution to
$\langle\rho+P\rangle$, which is proportional to $T^4$, is small compared to the
vacuum contribution from $\phi$. We will come back to this point at the end of this
subsection.

{}From the energy-momentum tensor for a scalar field, we find
\begin{align}
\rho & = T^0_{\,\,\,0} = \frac{1}{2}\dot\phi^2 +
\frac{1}{2}\frac{(\nabla\phi)^2}{a^2} + V(\phi) \, ,
\\
p & = -\frac{1}{3}T^i_{\,\,\,i} = \frac{1}{2}\dot\phi^2 -
\frac{1}{6}\frac{(\nabla\phi)^2}{a^2} - V(\phi) \, .
\end{align}
Thus we have
\begin{equation}
\rho + p = \dot\phi^2 + \frac{1}{3}\frac{(\nabla\phi)^2}{a^2} \, .
\end{equation}
Using the standard Fourier mode expansion of the scalar field,
\begin{equation}
\phi(x) = \int \frac{d^3k}{(2\pi)^{3/2}} \left[ a_{\bm{k}} \phi_k(\eta)
e^{i\bm{k\cdot x}}
 + a_{\bm{k}}^\dagger \phi_k^*(\eta) e^{-i\bm{k\cdot x}}\right] \, ,
\end{equation}
we then obtain
\begin{equation}
\langle \dot\phi^2 \rangle = \left\langle \frac{\phi'^2}{a^2} \right\rangle =
\frac{1}{a^2} \int \frac{d^3k}{(2\pi)^3} \phi_k'(\eta) {\phi_k^*}'(\eta) \,.
\end{equation}
The explicit form of the mode function $\phi_k$ and the calculation of $\langle
\dot\phi^2 \rangle$ are given in Appendix~\ref{modefcn}. We are left with
\begin{align}
\langle \dot\phi^2 \rangle & = \frac{\pi}{4a^3}H\left( \frac{9}{4} - \nu^2 \right)
\frac{4\pi}{(2\pi)^3} \int_0^\infty dk \, k^2 H_{\nu}^{(1)} (z) H_{\nu}^{(2)} (z)
\nonumber\\
& \hspace{0.5cm} + \frac{\pi}{4a^4}(-\eta)\left( \frac{1}{2} + \frac{3}{4\nu}
\right)\frac{4\pi}{(2\pi)^3} \int_0^\infty dk \, k^2 (k^2 + \mu^2)
H_{\nu-1}^{(1)} (z) H_{\nu-1}^{(2)} (z)
\nonumber\\
& \hspace{0.5cm} + \frac{\pi}{4a^4}(-\eta)\left( \frac{1}{2} - \frac{3}{4\nu}
\right)\frac{4\pi}{(2\pi)^3} \int_0^\infty dk \, k^2 (k^2 + \mu^2)
H_{\nu+1}^{(1)} (z) H_{\nu+1}^{(2)} (z) \, .
\end{align}
In the same way, we find
\begin{equation}
\frac{1}{3}\frac{(\nabla\phi)^2}{a^2} =
\frac{\pi}{12a^4}(-\eta)\frac{4\pi}{(2\pi)^3} \int_0^\infty dk \, k^4
H_\nu^{(1)}(z) H_\nu^{(2)}(z) \, .
\end{equation}
Now we have all the ingredients to calculate $\langle \rho + p \rangle$.

As clear from the above expressions, the momentum integrals diverge at large $k$.
This is the standard ultraviolet divergence that should be regularized or
renormalized. To regularze the divergence, we introduce a simple cutoff at a large
physical momentum, $k/a\leq (k/a)_c=H\Lambda$ where $\Lambda$ is a large number.

Now let us evaluate the integrals in the limit $\eta \rightarrow -\infty$, i.e.
during the early stage of inflation. This is the region of our interest. In this
case, we have
\begin{equation}
\sqrt{k^2 + \mu^2}(-\eta) > \mu(-\eta) \gg 1 \, ,
\end{equation}
so the Hankel functions are approximated as
\begin{equation}
H_\nu^{(1,2)}(z) \sim
 \sqrt{\frac{2}{\pi z}} e^{\pm i \left( z -
\frac{\pi}{2}\nu - \frac{\pi}{4} \right)} \quad
(z\gg1)\, ,
\end{equation}
thus
\begin{equation}
H_\nu^{(1)}(z) H_\nu^{(2)}(z) \sim \frac{2}{\pi z}
\quad(z\gg1) \, ,
\end{equation}
which is independent of the parameter $\nu$. Then after some calculations, we find
\begin{equation}
\langle \rho + p \rangle \, \underset{\Lambda \gg 1}{\longrightarrow} \,
\frac{H^2}{(2\pi)^2} \left\{ \frac{\Lambda^4}{3} +\left( \frac{9}{4} - \nu^2 \right)
\left[ \frac{\Lambda^2}{2} + \frac{\mu^2|\eta|^2}{4} \left( 1 -
\frac{\mu^2|\eta|^2}{2\Lambda^2} \right) + \frac{\mu^2|\eta|^2}{2}
\log\left(\frac{\mu|\eta|}{2\Lambda} \right)^2 \right] \right\}.
\label{rplusp}
\end{equation}
To renormalize the above expression, we assume that the de Sitter invariance will be
(formally) unbroken if $\mu=0$, that is, if there is no preferred frame. Then the
renormalized expectation value is given by
\begin{eqnarray}
\langle\rho+p\rangle_{\rm ren}
=\lim_{\Lambda\gg1}\left[\langle\rho+p\rangle(\Lambda,\mu)
-\langle\rho+p\rangle(\Lambda,0)\right]\,,
\end{eqnarray}
where $\langle\rho+p\rangle(\Lambda,\mu)$ is the vacuum expectation value given by
Eq.~(\ref{rplusp}). This leads to
\begin{equation}\label{rho+p_limit}
\langle \rho + p \rangle_{\rm ren}
=\frac{H^4}{16\pi^2} \frac{m_\phi^2}{H^2} (\mu\eta)^2
 \left\{ 1 + \log \left[ \left(\frac{\mu\eta}{2\Lambda} \right)^4
 \right] \right\} \,.
\end{equation}
This should be valid for $1 \ll \mu|\eta| \ll \Lambda$.

Now, let us check if the above result is indeed greater than the thermal
contribution. Assuming there are $N$ effective massless degrees of freedom that are
thermal and that couple to the $\phi$ field, we have an estimate
\begin{eqnarray}
(\rho+P)_{T}\sim N T^4
\quad \mbox{and}\quad \frac{\mu^2}{a^2}=g^2NT^2\,,
\end{eqnarray}
where the suffix $T$ stands for thermal, and we have assumed the same coupling
constant $g^2$ for all the thermal fields for simplicity. On the other hand, the
contribution from the vacuum fluctuations of $\phi$, Eq.~(\ref{rho+p_limit}), may be
approximately expressed as
\begin{eqnarray}
\langle \rho + p \rangle_{\phi}\sim \left|m_\phi^2\right|\,\frac{\mu^2}{a^2}
=\left|m_\phi^2\right|\,g^2NT^2\,.
\end{eqnarray}
Hence in order for the $\phi$ contribution to dominate, we must have $T^2\ll
\left|m_\phi^2\right|g^2$. Since the stage of our interest is
$\left|m_\phi^2\right|<\mu^2/a^2=g^2 NT^2$, this leads to the condition,
\begin{eqnarray}
T^2\ll \left|m_\phi^2\right|\,g^2<g^4N T^2\,.
\end{eqnarray}
This can be satisfied only if we have $g^4N\gg1$. Namely, the dominance of the
$\phi$ contribution over thermal corrections can be realized if there are
sufficiently large massless degrees of freedom which are in thermal equilibrium at
an early stage of the false vacuum inflation. However, we should mention that the
condition $g^4N\gg 1$ implies that the theory is in a regime of strong coupling.
Thus the thermal contribution can dominate in general, suggesting that thermal
corrections may be worth investigating in more detail, although it is outside of the
scope of the present paper. In the following subsection we assume that thermal
corrections are negligible.

\subsubsection{Extraction of growing solution}

Using the results obtained until now, we can compute the comoving curvature
perturbation $\mathcal{R}_c$. First let us recapitulate the standard result.

On super-horizon scales, for a completely general equation of state, we have the
general solution for $\Phi$,
\begin{equation}\label{Phigeneralsol}
\Phi = \frac{3}{2}C_1 \frac{\mathcal{H}}{a^2}
 \int_{\eta_i}^\eta (1 + w) a^2(\eta')d\eta' + \mbox{decaying mode} \, ,
\end{equation}
where $\mathcal{H} \equiv a'/a$, $w=p/\rho$ and the decaying mode is proportional to
$\mathcal{H}/a^2$. The initial time $\eta_i$ can be chosen arbitrarily since its
change only affects the contribution of the decaying mode.

Given the Newtonian curvature perturbation $\Phi$, $\mathcal{R}_c$ is expressed in
terms of $\Phi$ as
\begin{eqnarray}
\mathcal{R}_c=\frac{2\Phi'+(5+3w)\mathcal{H}\Phi}{3(1+w)\mathcal{H}}\,.
\label{calRexp}
\end{eqnarray}
Plugging the general solution given by Eq.~(\ref{Phigeneralsol}) into the above, we
find, neglecting the decaying mode part,
\begin{equation}
\mathcal{R}_c = C_1 \, .
\end{equation}
Thus the coefficient $C_1$ indeed corresponds to the amplitude of the growing
adiabatic mode.

In order to extract out the final curvature perturbation amplitude $C_1$ from our
calculation, we need to perform the integral in Eq.~(\ref{Phigeneralsol})
explicitly. To do so, we need the information of $a$ and $1+w$. At leading order
approximation, we may assume the universe is de Sitter. Hence we may approximate the
scale factor and the Hubble parameter as
\begin{align}
a = \frac{1}{-H\eta} \, ,\quad
\mathcal{H} = \frac{1}{-\eta} \, .
\end{align}
As for $1 + w$, however, we must take into account the small deviations from de
Sitter. From Eq.~(\ref{rho+p_limit}), we note that the logarithm $\log [
(\mu\eta/2\Lambda)^2 ]$ is negative for $\mu|\eta| \ll \Lambda$. So, neglecting the
slow logarithmic behavior and take into account the fact $m_\phi^2 < 0$, we
effectively have
\begin{equation}\label{rho+p_limit2}
\langle \rho + p \rangle_{\rm ren}
= A \frac{H^4}{16\pi^2} \frac{|m_\phi^2|}{H^2} (\mu\eta)^2\, ,
\end{equation}
where $A$ is $\mathcal{O}(1)$ and positive. This gives
\begin{eqnarray}
1+w=\frac{A}{48\pi^2}\frac{|m_\phi^2|}{\mpl^2}(\mu\eta)^2\,.
\end{eqnarray}
This should be valid for $\mu^2\eta^2\gg1$ as long as $1+w\lesssim O(1)$, that is,
\begin{eqnarray}
\frac{\mpl^2}{|m_\phi^2|}\gg\mu^2\eta^2\gg1\,.
\end{eqnarray}
Keeping in mind this range, the integral on the right-hand side of
Eq.~(\ref{Phigeneralsol}) can be easily evaluated to give
\begin{eqnarray}
\Phi=-C_1\kappa(\mu\eta)^2+\mbox{decaying mode}\,;
\quad
\kappa=\frac{A}{32\pi^2}\frac{|m_\phi^2|}{\mpl^2}\,.
\label{kappadef}
\end{eqnarray}

Now it is easy to obtain the final amplitude of $\mathcal{R}_c$. We just have to
divide the early time solution for $\Phi$ by $\kappa (\mu\eta)^2$. Then the spectrum
of the conserved comoving curvature perturbation is
\begin{equation}
\mathcal{P}_\mathcal{R}(k)
= \frac{\mathcal{P}_\Phi(k;\eta)}{\kappa^2(\mu\eta)^4} \,.
\end{equation}
Inserting Eq.~(\ref{Phispectrum}) into the above, we thus finally find the spectrum
of the comoving curvature perturbation at $k\ll\mu$ as
\begin{equation}\label{curvspec}
\mathcal{P}_\mathcal{R}(k) \approx \frac{B}{A^2} \left( \frac{H^2}{m_\phi^2}
\right)^2 \left( \frac{k}{\mu} \right)^3 \, ,
\end{equation}
where $B=-2{\rm Ei(-2)}\approx 0.0978010$. Thus, with $\mathcal{P}_\mathcal{R}(k)
\propto k^{n_\mathcal{R} - 1}$, where $n_\mathcal{R}$ is the spectral index, we have
\begin{equation}\label{curvspecindex}
n_\mathcal{R} = 4 \, ,
\end{equation}
i.e. a very blue spectrum. This result is in accordance with a naive expectation.
That is, for a very small $k$ which leaves the horizon when $\phi=0$ is strongly
stable, the generation of the curvature perturbation must be severely suppressed. On
the other hand, as $k$ approaches $\mu$, the amplitude of the curvature perturbation
should increase. This is because the effect of the negative $m_\phi^2$ gradually
becomes more and more important as the universe expands and the instability sets in
at
\begin{eqnarray}
\mu^2\eta^2\leq\mu^2\eta_c^2=\frac{|m_\phi^2|}{H^2}\,.
\end{eqnarray}
It is then expected that the fluctuations would become very large at $\eta=\eta_c$.

\section{Conclusions}
\label{section_conclusions}

In this paper, we have calculated the power spectrum $\mathcal{P}_\mathcal{R}$ and
the corresponding spectral index $n_\mathcal{R}$ of the comoving curvature
perturbation $\mathcal{R}_c$ produced when the inflaton field is trapped in a local
minimum of the effective potential with non-zero vacuum energy. The difficulty we
confront when we adopt the conventional approach of the calculation of
$\mathcal{R}_c$ is that classically the inflaton is well anchored if
$|m_\mathrm{eff}| \gg 3H/2$ so that $\dot\phi = 0$, and accordingly comoving
hypersurfaces on which $\mathcal{R}_c$ is given become singular: this is because we
cannot define comoving hypersurfaces when $\dot\phi = 0$.

To evade this difficulty, we have used a pure quantum field theory approach to
calculate the two-point correlation function of the inflaton field and the
perturbation of the energy density in the exact de Sitter background. This de Sitter
phase, due to a non-vanishing false vacuum energy, shoud not last forever but should
eventually be terminated so that the standard hot big bang evolution of the universe
can commence. We achieve this by adding a comoving mass term, or equivalently a
thermal correction, $\mu^2/a^2$, to the potential as shown in Eq.~(\ref{mass}). This
breaks the perfect de Sitter invariance, and allows the vacuum expectation value of
$\rho+p$ to be non-vanishing. Then given the fact that $\langle\rho+p\rangle\neq0$,
we have explicitly calculated the final comoving curvature power spectrum
$\mathcal{P}_\mathcal{R}$, given by Eq.~(\ref{curvspec}). The spectral index is
found to be very blue, $n_\mathcal{R}=4$.

We believe our results are widely applicable: for example, we can directly
constrain the production of the primordial black holes after thermal
inflation~\cite{hopefully}, which has not yet been studied anywhere including the
original references~\cite{thermalinflation}. Also since the curvature perturbation
is quadratic in the scalar field it is highly non-Gaussian. This issue will also be
reported separately.

\subsection*{Acknowledgements}

We thank Daniel Chung, Dmitry Gal'tsov, Jai-chan Hwang, Nemanja Kaloper, Andrei
Linde, Thanu Padmanabhan, Ewan Stewart and Takahiro Tanaka for helpful discussions.
 JG is also grateful to the Yukawa Institute for Theoretical Physics at
Kyoto University where some part of this work was carried out during ``Scientific
Program on Gravity and Cosmology'' (YITP-T-07-01) and ``KIAS-YITP Joint Workshop:
String Phenomenology and Cosmology'' (YITP-T-07-10).
 JG is partly supported by the Korea Research Foundation Grant KRF-2007-357-C00014
funded by the Korean Government.
MS is supported in part by JSPS Grant-in-Aid for
Scientific Research (B) No.~17340075 and (A) No.~18204024,
and by JSPS Grant-in-Aid for Creative Scientific Research No.~19GS0219.

\appendix

\section{Inflaton field mode functions}
\label{modefcn}

Since there is no classically evolving background, the quantization of the inflaton
can be done without worrying about the metric perturbation, that is, as in the
standard quantization of a scalar field in curved spacetime.

We start with the Fourier expansion of the inflaton field,
\begin{equation}
\phi(x) = \int \frac{d^3k}{(2\pi)^{3/2}} \left[ a_{\bm{k}} \phi_k(\eta)
e^{i\bm{k\cdot x}}
 + a_{\bm{k}}^\dagger \phi_k^*(\eta) e^{-i\bm{k\cdot x}}\right] \, ,
\end{equation}
where the annihilation and creation operators $a_{\bm{k}}$ and $a_{\bm{k}}^\dagger$
satisfy the standard commutation relation
\begin{equation}
\left[ a_{\bm{k}}, a_{\bm{q}}^\dagger \right]
= \delta^{(3)}(\bm{k}-\bm{q}) \, .
\end{equation}
The mode function $\phi_k$ is called the positive frequency function. It determines
the vacuum annihilated by $a_{\bm{k}}$. The mode equation is
\begin{eqnarray}
\phi_k'' - \frac{2}{\eta}\phi_k' + \left[ \left( k^2 + \mu^2 \right) +
\frac{m_\phi^2}{H^2\eta^2} \right]\phi_k = 0 \, ,
\end{eqnarray}
with the normalization
\begin{eqnarray}
\phi_k'\phi_k^*-\phi_k^{*}{}'\phi_k=-\frac{i}{a^2}=-iH^2\eta^2\,,
\end{eqnarray}
where a prime denotes a derivative with respect to $\eta$.

The positive frequency function $\phi_{\bm{k}}$ appropriate for the inflationary
universe is given by
\begin{equation}\label{solphi}
\phi_k(\eta) = \frac{\sqrt{\pi}}{2}H(-\eta)^{3/2}H_\nu^{(1)}\left( -\sqrt{k^2 +
\mu^2}\eta \right) \, ,
\end{equation}
where $\nu$ is given by Eq.~(\ref{nu}),
\begin{eqnarray}
\nu=\sqrt{\frac{9}{4}-\frac{m_\phi^2}{H^2}}
~\left(>\frac{3}{2}\right),
\end{eqnarray}
and $H_\nu^{(1)}(z)$ is the Hankel function of the first kind. This mode function
satisfies the asymptotic boundary condition that it reduces to the one for Minkowski
vacuum in the limit $\eta\to-\infty$, corresponding to the high frequency limit
where the cosmic expansion can be totally neglected. Below, we list a few formulas
for $\phi_k$ and its derivatives which are used for the computation of the
components of the energy-momentum tensor.

Taking a time derivative of Eq.~(\ref{solphi}), we find
\begin{align}
\phi_k'(\eta) & = \frac{d}{d\eta} \left[
\frac{\sqrt{\pi}}{2}H(-\eta)^{3/2}H_\nu^{(1)}\left( -\sqrt{k^2 + \mu^2}\eta \right)
\right]
\nonumber \\
& =\frac{\sqrt{\pi}}{2}H(-\eta)^{1/2}
\left[-\frac{3}{2} H_\nu^{(1)}\left( -\sqrt{k^2 +
\mu^2}\eta \right) - \sqrt{k^2 + \mu^2}(-\eta)
{H_\nu^{(1)}}'\left( -\sqrt{k^2 + \mu^2}\eta
\right)\right] \, ,
\end{align}
where the prime of the Hankel function means the derivative with respect to the
whole argument, i.e. ${H_\nu^{(1)}}'(z) = dH_\nu^{(1)}(z)/dz$.

Using ${H_\nu^{(1)}}^*(z) = H_\nu^{(2)}(z)$ for real $\nu$ and $z$, we find
\begin{align}
\phi_k'(\eta){\phi_k^*}'(\eta) & = \frac{\pi}{4}H^2(-\eta) \left\{
\frac{9}{4}H_\nu^{(1)}(z)H_\nu^{(2)}(z) + (k^2 +
\mu^2)(-\eta)^2{H_\nu^{(1)}}'(z){H_\nu^{(2)}}'(z) \right.
\nonumber \\
& \hspace{2.5cm} \left. + \frac{3}{2}\sqrt{k^2 + \mu^2}(-\eta) \left[
H_\nu^{(1)}(z){H_\nu^{(2)}}'(z) + {H_\nu^{(1)}}'(z)H_\nu^{(2)}(z) \right] \right\}
\,,
\end{align}
where $z=\sqrt{k^2+\mu^2}\,(-\eta)$. Eliminating $H_\nu'(z)$ by using the Hankel
function identities
\begin{align}
H_\nu'(z) = & \frac{H_{\nu-1}(z) - H_{\nu+1}(z)}{2} \, ,
\\
H_\nu'(z) = & \frac{\nu}{z}H_\nu(z) - H_{\nu+1}(z) \, ,
\\
H_\nu(z) = & \frac{z}{2\nu} \left[ H_{\nu-1}(z) + H_{\nu+1}(z) \right] \, ,
\end{align}
which hold for both $H_\nu^{(1)}(z)$ and $H_\nu^{(2)}(z)$, we obtain
\begin{align}
\phi_k'(\eta){\phi_k^*}'(\eta)
=& \frac{\pi}{4}H^2(-\eta) \left[ \left( \frac{9}{4}
- \nu^2 \right) H_\nu^{(1)} (z) H_\nu^{(2)} (z) \right.
\nonumber\\
&\left.\hspace{2cm} + \left(\frac{1}{2} + \frac{3}{4\nu} \right)
z^2H_{\nu - 1}^{(1)} (z) H_{\nu - 1}^{(2)} (z)
+ \left(\frac{1}{2} - \frac{3}{4\nu} \right)
 z^2 H_{\nu + 1}^{(1)} (z) H_{\nu + 1}^{(2)} (z) \right] \, .
\label{dphidphi}
\end{align}

\section{Formulas for energy density two-point function}
\label{2pointfct_formula}

Here we give an explicit expression for the two-point function $D(x,x')$ introduced
in Eq.~(\ref{corrD}) in terms of the scalar field two-point function $G(x,x')$. We
also give useful formulas for the spatial and time derivatives when they act on
$G(x,x')$.

Substituting the expressions for the coefficients $f^{\rho\mu\nu}_j$ given by
Eqs.~(\ref{coeff1}), (\ref{coeff2}) and (\ref{coeff3}) into Eq.~(\ref{corrD}), we
find
\begin{align}\label{expD}
D(x, x') & = \partial^i\partial^{j'} \left[ \left(
\partial_i\partial_0\partial_{0'}G \right) \left(
\partial_{j'}\partial_0\partial_{0'}G \right) + \left(
\partial_0\partial_{0'}G \right) \left(
\partial_i\partial_{j'}\partial_0\partial_{0'}G \right) \right]
\nonumber \\
& \hspace{1.5cm} - \left( \partial_0\partial_{0'}^2G \right) \left(
\partial_i\partial_{j'}\partial_0G \right) - \left(
\partial_{j'}\partial_0 G \right) \left( \partial_i\partial_0\partial_{0'}^2G
\right)
\nonumber \\
& \hspace{1.5cm} - \left( \partial_i\partial_{j'}\partial_{0'}G \right) \left(
\partial_{0'}\partial_0^2 G \right) - \left( \partial_i\partial_{0'}G \right) \left(
\partial_{j'}\partial_{0'}\partial_0^2G \right)
\nonumber \\
& \hspace{1.5cm} \left. + \left( \partial_i\partial_{0'}^2G \right) \left(
\partial_{j'}\partial_0^2G \right) + \left( \partial_i\partial_{j'}G \right) \left(
\partial_0^2\partial_{0'}^2G \right) \right]
\nonumber \\
& \hspace{.5cm} + a^{-2} \left\{ \partial^i\partial^{j'} \left[ \left(
\partial^{k'}\partial_{k'}\partial_0 G \right) \left(
\partial_i\partial_{j'}\partial_0 G \right) + \left( \partial_{j'}\partial_0 G
\right) \left( \partial_i \partial^{k'}\partial_{k'}\partial_0 G \right) \right.
\right.
\nonumber \\
& \hspace{3cm} + \left( \partial_{j'}\partial^{k'}\partial_0 G \right) \left(
\partial_i\partial_{k'}\partial_0 G \right) + \left( \partial_{k'}\partial_0 G
\right) \left( \partial_i\partial_{j'}\partial^{k'}\partial_0 G \right)
\nonumber \\
& \hspace{3cm} - \left( \partial_i\partial^{k'}\partial_{k'} G \right) \left(
\partial_{j'}\partial_0^2 G \right) - \left( \partial_i\partial_{j'} G \right)
\left( \partial^{k'}\partial_{k'}\partial_0^2 G \right)
\nonumber \\
& \hspace{3cm} - \left(\partial_i\partial_{j'}\partial^{k'} G \right) \left(
\partial_{k'}\partial_0^2 G \right) - \left( \partial_i\partial_{k'} G \right)
\left( \partial_{j'}\partial^{k'}\partial_0^2 G \right)
\nonumber \\
& \hspace{3cm} + \left( \partial_i\partial_{j'}\partial_{0'} G \right) \left(
\partial_k\partial^k\partial_{0'} G \right) + \left( \partial_i\partial_{0'} G
\right) \left( \partial_k\partial^k\partial_{j'}\partial_{0'} G \right)
\nonumber \\
& \hspace{3cm} + \left( \partial_k\partial_{j'}\partial_{0'} G \right) \left(
\partial_i\partial^k\partial_{0'} G \right) + \left( \partial_k\partial_{0'} G
\right) \left( \partial_i\partial^k\partial_{j'}\partial_{0'} G \right)
\nonumber \\
& \hspace{3cm} - \left( \partial_i\partial_{0'}^2 G \right) \left(
\partial_k\partial^k\partial_{j'} G \right) - \left( \partial_i\partial_{j'} G
\right) \left( \partial_k\partial^k\partial_{0'}^2 G \right)
\nonumber \\
& \hspace{3cm} \left. \left. - \left( \partial_k\partial_{0'}^2 G \right) \left(
\partial_i\partial^k\partial_{j'} G \right) - \left( \partial_k\partial_{j'} G
\right) \left( \partial_i\partial^k\partial_{0'}^2 G \right) \right] \right\}
\nonumber \\
& \hspace{.5cm} + a^{-4} \left\{ \partial^i\partial^{j'} \left[ \left(
\partial_i\partial_{l'}\partial^{l'} G \right) \left(
\partial_k\partial^k\partial_{j'} G \right) + \left( \partial_i\partial_{j'} G
\right) \left( \partial_k\partial^k\partial_{l'}\partial^{l'} G \right) \right.
\right.
\nonumber \\
& \hspace{3cm} + \left( \partial_i\partial_{j'}\partial^{l'} G \right) \left(
\partial_k\partial^k\partial_{l'} G \right) + \left( \partial_i\partial_{l'} G
\right) \left( \partial_k\partial^k\partial_{j'}\partial^{l'} G \right)
\nonumber \\
& \hspace{3cm} + \left( \partial_k\partial_{l'}\partial^{l'} G \right) \left(
\partial_i\partial^k\partial_{j'} G \right) + \left( \partial_k\partial_{j'} G
\right) \left( \partial_i\partial^k\partial_{l'}\partial^{l'} G \right)
\nonumber \\
& \hspace{3cm} \left. \left. + \left( \partial_k\partial_{j'}\partial^{l'} G \right)
\left( \partial_i\partial^k\partial_{l'} G \right) + \left(
\partial_k\partial_{l'} G \right) \left(
\partial_i\partial^k\partial_{j'}\partial^{l'} G \right) \right] \right\} \, .
\end{align}
Note that we have not used any special properties of $G(x,x')$ in the above.

To proceed, we first make use of the properties of $G(x,x')$. As far as the spatial
dependence is concerned, it depends only on the distance between two points $r =
|\bm{x} -\bm{x}'|$. Hence we have
\begin{equation}\label{xx'deriv}
\partial_{x'} = -\partial_x \, .
\end{equation}
As for the time dependence, $G(x,x')$ is symmetric in the interchange of $t$ and
$t'$. Thus we have
\begin{eqnarray}
G=G(r;t,t')=G(r;t',t)\,.
\label{Gprop}
\end{eqnarray}

Now, for the two-point function $G(x,x')$ with the above properties, the spatial
derivatives acting on $G(x,x')$ may be expressed as
\begin{align}\label{spderiv}
\partial_i & = \frac{x^i}{r}\partial_r \, ,
\nonumber\\
\partial_i\partial_j & = \frac{x^ix^j}{r^2}\partial_r^2 + \left( \delta_{ij} -
\frac{x^ix^j}{r^2} \right) \frac{1}{r}\partial_r \, ,
\nonumber\\
\partial_i\partial_j\partial_k & = \frac{x^ix^jx^k}{r^3}\partial_r^3 +
\left( \delta_{ij}\frac{x^k}{r} + \delta_{jk}\frac{x^i}{r} +
\delta_{ki}\frac{x^j}{r} - \frac{3x^ix^jx^k}{r^3} \right) \left(
\frac{1}{r}\partial_r^2 - \frac{1}{r^2}\partial_r \right) \, ,
\nonumber\\
\partial_i\partial_j\partial_k\partial_l & = \frac{x^ix^jx^kx^l}{r^4}\partial_r^4
\nonumber\\
& \hspace{.5cm} + \left( \delta_{ij}\frac{x^kx^l}{r^2} +
\delta_{ik}\frac{x^jx^l}{r^2} + \delta_{il}\frac{x^jx^k}{r^2} +
\delta_{jk}\frac{x^ix^l}{r^2} + \delta_{jl}\frac{x^ix^k}{r^2} +
\delta_{kl}\frac{x^ix^j}{r^2} - \frac{6x^ix^jx^kx^l}{r^4} \right)
\frac{1}{r}\partial_r^3
\nonumber\\
& \hspace{.5cm} + \left[ \delta_{ij}\delta_{kl} + \delta_{jk}\delta_{li} +
\delta_{ki}\delta_{jl} - 3 \left( \delta_{ij}\frac{x^kx^l}{r^2} +
\delta_{ik}\frac{x^jx^l}{r^2} + \delta_{il}\frac{x^jx^k}{r^2} +
\delta_{jk}\frac{x^ix^l}{r^2} + \delta_{jl}\frac{x^ix^k}{r^2} +
\delta_{kl}\frac{x^ix^j}{r^2} \right) \right.
\nonumber\\
& \hspace{1.2cm} \left. + \frac{15x^ix^jx^kx^l}{r^4} \right] \left(
\frac{1}{r^2}\partial_r^2 - \frac{1}{r^3}\partial_r \right) \, .
\end{align}
Similarly, for the time derivatives acting on $G(x,x')$,
using $d\eta = dt/a$, we have
\begin{align}\label{tderiv}
\partial_0 & = H(-\eta)\partial_\eta \, ,
\nonumber\\
\partial_0^2 &
= H^2(-\eta) \left[ (-\eta)\partial_\eta^2 - \partial_\eta \right] \,,
\nonumber\\
\partial_0\partial_{0'} & = H^2(\eta\eta') \partial_\eta\partial_{\eta'} \, ,
\nonumber\\
\partial_{0'}\partial_0^2 & = H^3(\eta\eta') \left[
(-\eta)\partial_{\eta'}\partial_\eta^2 - \partial_{\eta'}\partial_\eta \right]\,,
\nonumber\\
\partial_0^2\partial_{0'}^2 & = H^4 (\eta\eta') \left[
(\eta\eta')\partial_\eta^2\partial_{\eta'}^2 -
(-\eta)\partial_\eta^2\partial_{\eta'}
- (-\eta')\partial_\eta\partial_{\eta'}^2+\partial_\eta\partial_{\eta'}\right]\,.
\end{align}
These formulas are used in the explicit evaluation of $D(x,x')$.


\begin{thebibliography}{999}


\bibitem{Guth:1980zm}
  A.~H.~Guth,
  Phys.\ Rev.\  D {\bf 23}, 347 (1981).


\bibitem{inf}
  A.~D.~Linde,
  Phys.\ Lett.\  B {\bf 108}, 389 (1982)~;
  A.~Albrecht and P.~J.~Steinhardt,
  Phys.\ Rev.\ Lett.\  {\bf 48}, 1220 (1982).


\bibitem{Liddle:2000cg}
  A.~R.~Liddle and D.~H.~Lyth,
  ``Cosmological inflation and large-scale structure,''
{\it  Cambridge, UK: Univ. Pr. (2000) 400 p}


\bibitem{Komatsu:2008hk}
  E.~Komatsu {\it et al.}  [WMAP Collaboration],
  arXiv:0803.0547 [astro-ph].


\bibitem{Kodama:1985bj}
  H.~Kodama and M.~Sasaki,
  Prog.\ Theor.\ Phys.\ Suppl.\  {\bf 78}, 1 (1984).


\bibitem{Mukhanov:1990me}
  V.~F.~Mukhanov, H.~A.~Feldman and R.~H.~Brandenberger,
  Phys.\ Rept.\  {\bf 215}, 203 (1992).


\bibitem{higherordercalc}
  J.~O.~Gong and E.~D.~Stewart,
  Phys.\ Lett.\  B {\bf 510}, 1 (2001)
  [arXiv:astro-ph/0101225]~;
  J.~Choe, J.~O.~Gong and E.~D.~Stewart,
  JCAP {\bf 0407}, 012 (2004)
  [arXiv:hep-ph/0405155].


\bibitem{Dvali:2003vv}
  G.~Dvali and S.~Kachru,
  arXiv:hep-th/0309095.


\bibitem{thermalinflation}
  D.~H.~Lyth and E.~D.~Stewart,
  Phys.\ Rev.\ Lett.\  {\bf 75}, 201 (1995)
  [arXiv:hep-ph/9502417]~;
  D.~H.~Lyth and E.~D.~Stewart,
  Phys.\ Rev.\  D {\bf 53}, 1784 (1996)
  [arXiv:hep-ph/9510204].


\bibitem{Gong:2006hf}
  See, e.g. J.~O.~Gong,
  Phys.\ Lett.\  B {\bf 637}, 149 (2006)
  [arXiv:hep-ph/0602106].


\bibitem{Pilo:2004ke}
  L.~Pilo, A.~Riotto and A.~Zaffaroni,
  Phys.\ Rev.\ Lett.\  {\bf 92}, 201303 (2004)
  [arXiv:astro-ph/0401302].


\bibitem{Nambu:1989eu}
  Y.~Nambu and M.~Sasaki,
  Prog.\ Theor.\ Phys.\  {\bf 83}, 37 (1990).


\bibitem{Linde:1982uu}
  A.~D.~Linde,
  Phys.\ Lett.\  B {\bf 116}, 335 (1982).


\bibitem{Bunch:1978yq}
  T.~S.~Bunch and P.~C.~W.~Davies,
  Proc.\ Roy.\ Soc.\ Lond.\  A {\bf 360}, 117 (1978).


\bibitem{hopefully}
  D.~J.~H.~Chung, J.~O.~Gong and M.~Sasaki, in preparation


\end{thebibliography}
\end{document}